# WNtags: A Web-Based Tool For Image Labeling And Retrieval With Lexical Ontologies

Marko Horvat, Anton Grbin, Gordan Gledec

Faculty of Electrical Engineering and Computing, University of Zagreb,
Unska 3, HR-10000 Zagreb, Croatia
`{Marko.Horvat2, Anton Grbin, Gordan.Gledec}@fer.hr`

**Abstract.** Ever growing number of image documents available on the Internet continuously motivates research in better annotation models and more efficient retrieval methods. Formal knowledge representation of objects and events in pictures, their interaction as well as context complexity becomes no longer an option for a quality image repository, but a necessity. We present an ontology-based online image annotation tool WNtags and demonstrate its usefulness in several typical multimedia retrieval tasks using International Affective Picture System emotionally annotated image database. WNtags is built around Word-Net lexical ontology but considers Suggested Upper Merged Ontology as the preferred labeling formalism. WNtags uses sets of weighted WordNet synsets as high-level image semantic descriptors and query matching is performed with word stemming and node distance metrics. We also elaborate our near future plans to expand image content description with induced affect as in stimuli for research of human emotion and attention.

## 1 Introduction

Image repositories annotated with ground truth labels are welcome tools in many areas of computer science which address the problems of visual data classification, recognition or retrieval. Prevalence of visual media on the Web stresses practical uses of such repositories. Image retrieval algorithms with high precision and recall, and low fall-out are being increasingly sought as the number of indexed pictures on the Internet is getting larger. But because building a large dataset of annotated images is a costly and lengthy endeavor, it is important to create a flexible and modular system with small footprint that reuses available off the shelve components as much as possible. These motives guided planning and implementation of the WNtags ontology-based image annotation tool.

The WNtags allows image tagging with weighted sets of concepts. In the current version WNtags uses WordNet (hence the tool's name) as the knowledge database but its system architecture can accommodate concepts from other truly formal ontologies as image semantics descriptors. Apart from providing user friendly Web interface for collaborative image annotation, WNtags is also a tool for searching emotionally annotated real-life images. The image retrieval is based on semantic similarity between the search query and stored image annotations. This is accomplished using circumplex model of emotions [1], ontology-based image labels [2], corpus independent graph-based distance metrics [3] which, as the entire tool, has modular implementation and can be optimized or replaced independently from other parts of the system.

WNtags tool is different from other similar tools for manual image annotation (per example see [4][5]) in several respects. Firstly, WNtags uses International Affective Picture System (IAPS) dataset with semantically and emotionally described photographs that induce negative, neutral or positive affective reactions in stimulated persons [6]. IAPS pictures are annotated sparsely and inadequately using free-text keywords [7]. WNtags can also be viewed as a tool that significantly improves functionality and usability of this and other similar databases. Secondly, WNtags' modular construction test enables experimentation with different algorithms in various stages of image retrieval. Thirdly, WNtags uses WordNet [8] for description of image content and in the future envisages applying Suggested Upper Merged Ontology (SUMO) [9] to formally schematize image concepts. Finally, WNtags uses weighted measures and other heuristics to more accurately describe the most semantically important concepts in a picture and to reconcile different measures of observer variability.

The remainder of this paper is organized as follows. Chapter 2 discusses the WNtags as the collaborative web-based research tool in the area of image annotation, knowledge representation and document retrieval and describes its architecture. Chapter 3 describes formal methods of image annotation. Chapter 4 focuses on image retrieval and discusses the results. Finally, chapter 5 gives insight into future work and concludes the paper.

## 2  The WNtags collaborative web-based annotation tool

The WNtags is primarily a research tool in areas of image annotation, knowledge representation and multimedia document retrieval on the Web. Its development was motivated by a number of objectives and possible applications. In this section we describe the goals behind WNtags inception, its system architecture and the images used to test the tool's annotation and retrieval capabilities. There are two groups of motives or incentives for development of WNtags tool (Table 1). The goals are divided in two broad and complementary areas in computer science and artificial intelligence: knowledge representation and image retrieval.

**Table 1.** Goals motivating the WNtags tool development with possible applications.

| Goals | Possible applications |
| --- | --- |
| Knowledge representation | Image content description, annotation expressivity, semantic gap, affective dimensions, … |
| Image retrieval | Image repositories, emotionally annotated databases, web directories, … |

The most important goal of this project was to develop a collaborative online tool that could be modified and customized with minimal effort. In doing this, at least in the first version of WNtags, the system requirements were confined to representation of high-level semantics in static images. Some compromises had to be made since we wanted semantics annotations to be formal, but at the same close to a natural language. With the aid of a reasoning engine, the semantics annotations should be able to yield new knowledge about the image content. This is by no means an easy task, so we decided to use WordNet glossary to tag objects, events and affect in images. Our aim was to get an exhaustive set of relevant high-level semantics for each image in the database described with tags from a controlled glossary with graph-like structure. Such tag vocabulary can describe minimum necessary functional relationships between tags and discriminate among different meanings of the same lexical unit (i.e. word). Furthermore, we wanted all tags to be weighted to indicate their importance in aggregation of image content.

Representation of picture segments and objects is currently not in our focus as with some other image annotation tools [4] [5]. However, we plan to scale out the tool to be able to annotate and store lower level semantics as well. In the future we also want to add support for other multimedia formats like sounds or video-clips, but foremost for specific domains of static images such as facial expressions and especially affectively annotated images. Our goal is to optimally reuse the existing code base and gradually build up the application's features.

The secondary goal behind the WNtags project was experimentation in information retrieval using the described weighted high-level semantics from a controlled graph glossary. Instead of using statistical methods and bag-of-words concept in document indexing and retrieval, we would like to see images precisely annotated with least number of semantically meaningful tags. With formal logical systems based on Description Logic (DL) [10] these annotation may constitute a knowledge database of an intelligent expert system. Through the process of reasoning, the annotations could infer new knowledge about image content, effectively expanding the initial set of image tags. In this scheme, a search query given by a WNtags user becomes a search strategy goal in which the system must find the most closely matched images, fetch them from the database and sort the results by relevance. Procedures of user feedback and information refinement are also possible.

Finally, WNtags could also be interpreted as a step towards improvement of emotionally annotated databases such as IAPS [6], International Affective Digitized Sounds (IADS) [11] or Geneva Affective Picture Database (GAPED) [12]. These databases of multimedia stimuli are important for research of human emotion and attention [6], but also find their uses in a wide range of research (examples [13][14]). As mentioned before, their description schemes of multimedia semantics are currently quite rudimentary and limiting making image retrieval difficult and operator intensive [7]. Enhancement of multimedia content retrieval in these databases could substantially extend their use and promote further applications in cognitive sciences, psychology, psychiatry and neurology among others areas.

## 3 Formal image annotation

Optimally the annotation tool should not restrict the users in selection of text labels for describing images. Most users prefer to provide short and basic-level object labels (e.g. "child", "car", "person", "airplane"). This is not enough to establish a rich tag glossary which would enable quality image retrieval. At the same time, there can be a large variance of terms describing the same object category. Taken together, these make analysis and retrieval of the object labels difficult, since the system has to be aware of label synonyms, subsumed labels, member labels, and distinguish between object identity, events executed by objects, and attributes of objects and events.

We believe that only a knowledge database based on a large upper ontology and extended with appropriate domain ontologies can encapsulate semantics present in an image repository [15]. Naturally, domain ontologies would be selected depending on the repository's purpose and integrated with the basis upper ontology. Creating such image repository with large knowledge database and filling it with appropriately annotated images is an immense task. However, a compromise can be made with selection of WordNet as the tagging glossary.

WordNet is a lexical database of English language [8] which may also be interpreted as an informal and lexical ontology [16]. It is readily available, simple to use and – most importantly – contains over 100,000 concepts, i.e. synsets. This is ideal in terms of glossary size and encompasses any label a user may enter. However, WordNet's linguistic tokens lack the formality of ontological concepts. A formal ontology contains rules specifying complex relations that cannot be captured explicitly with simple links in graph-like knowledge structure. In spite of these shortcomings, WordNet is still far superior in image annotation than existing IAPS keywords, which are inconsistent, ambiguous and non-contiguous [7][17].

In WNtags, images are manually annotated with individual senses $s_1, s_2, \ldots, s_n$ of WordNet synsets $syn_1, syn_2, \ldots, syn_m$. Each image is tagged with at least three senses by a group of two or more individuals. Every sense $s_i$ is weighted in a range $w_i \in [0,1]$ according to its perceived relevance. Apart from weighted semantic labels, each image $img_i$ retained its original semantic description from IAPS $iaps_i$. IAPS keyword vocabulary $V^{IAPS}$ is unsupervised and each IAPS picture is described with a single free text keyword. Effectively, $iaps_i$ is one keyword from this corpus. IAPS uses circumplex model of affect to annotate stimuli emotion values [1][6]. In a nutshell, this model numerically describes the meaning of the emotional impulse in a 3-D space with the respect to the axes of valence or pleasure (denoted *val*), the axis of excitation or arousal (*ar*) and the axis of dominance over emotion (*dom*). Values of all axis are numerical and normalized in interval [1, 9].

Every image $img_i$ in WNtags repository retained its original IAPS emotional tuple value $emo_i$ with values $val_i$, $ar_i$ and $dom_i$. If $sem_i$ is set of all WordNet senses for $img_i$, $sêm_i$ set of weighted senses $sem_i$, and $iaps_i$ is one keyword from IAPS vocabulary then the cumulative semantic description $desc_i$ for one image $img_i$ stored in WNtags image repository from becomes a tuple

$$desc_i = \{\hat{sem}_i, emo_i, iaps_i\} \quad (1)$$

where

$$sem_i = (s_{i_1}, s_{i_2}, \ldots, s_{i_n})$$
$$\hat{sem}_i = (w_{i_1} s_{i_1}, w_{i_2} s_{i_2}, \ldots, w_{i_n} s_{i_n}) \quad (2)$$
$$emo_i = (val_i, ar_i, dom_i)$$
$$iaps_i \in V_i^{IAPS}$$

For every sense $s_i$ WNtags considers WordNet's built-in semantic relations hypernymy, hyponymy, holonymy and meronymy within a preset node distance $d$. WNtags also considers synonyms (coordinate terms) of $s_i$. Therefore for every sense $s_k$ in WordNet database and every sense $s_i$ in $desc_i$ broadens image description with all senses of all its neighboring synsets

$$sem_i = (s_{i_1}, s_{i_2}, \ldots, s_{i_n}) \cup \left( \bigcup_{s_j = s_{i_1}, s_{i_2}, \ldots, s_{i_n}} |(s_j, s_k)| \leq d \right) \quad (3)$$

As an example, IAPS annotated image (7175.jpg) with weighted labels and interaction with currently implemented WordNet knowledge database is shown in the next figure:

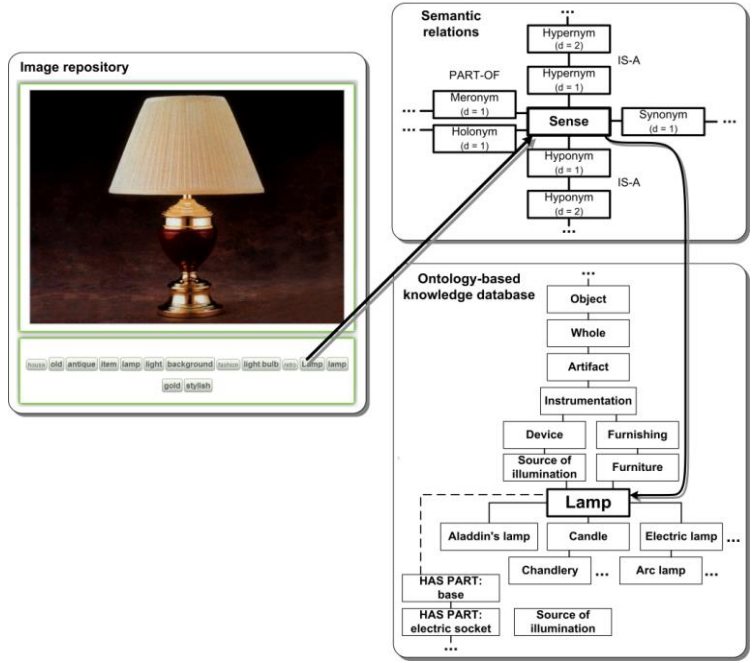

**Fig. 1.** An IAPS image in WNtags manually annotated with weighted concepts from a formal glossary and described in WordNet lexical ontology knowledge database. Each concept (e.g. "Lamp") has semantic neighborhood defined with semantic relations and node distance $d = 1, 2, 3, \ldots$

All indexed images are manually and collaboratively annotated. Personal bias of users will certainly result in different weights assigned to the image labels. Therefore, if there are $k$ weight ratings for $s_i$ they are averaged out $\bar{w}_i = \frac{1}{k}\sum_{j=1}^{k} w_{ij}$. Almost all users agreed on sets of 3-5 distinct labels per single image while finding additional descriptions in many cases proved difficult especially for semantically simple or abstract pictures.

In order to prove the validity of the WNtags concept we have randomly selected 100 emotionally annotated images with different object categories from IAPS database and tagged them using 956 different WordNet synsets. On average a single image is annotated with median of 20 WordNet tags (mean = 20.56436, sd = 2.76917, min=13, max=28) which were weighted according to their subjective importance in a particular image.

## 4 Image retrieval

The described WNtags annotating configuration allows multimodal image retrieval by three different information dimensions: affect, original free-text keywords and sets of synonyms from the WordNet lexical ontology. WNtags has modular architecture and in the current version, to save processing time, semantic similarity is loaded from the freely available WordNet::Similarity dataset with prepared similarity and relatedness pairs [18].

The search algorithm uses all senses of all collocations (i.e. non-permutable combinations) of query concepts. Some more complex multiword collocations may be found as specific senses in WordNet. All such detected query senses $qs_i$ in query $q_k$ are individually aligned with annotating senses $s_i$ for all images $img_l$, $l = 1, 2, \ldots, N_{img}$ where $N_{img}$ is the total number of images stored in the repository. Semantic distance between each pair is calculated together with subjective importance $\bar{w}_j$ of each $s_j$ in image $img_l$. Search goal function is

$$\max \sum_{\forall qs_i \in q_k, \forall s_j \in img_l} \bar{w}_j sim(qs_i, s_j) \quad (4)$$

An exhaustive search is executed and the retrieved results are sorted from the highest to the lowest relevance with possible values in range. The final results are rendered and displayed to the user.

To test the retrieval performance we performed semantic searches. All queries consisted of one WordNet concept randomly selected within $d = 30$ node distance between the nearest image tag. For example (as in Fig. 2.): "aircraft", "car", "boat", "helicopter", "road", "bus", etc.

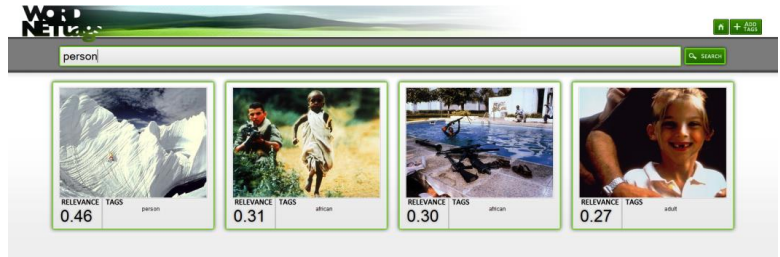

**Fig. 2.** Retrieved results from WNtags repository after querying it with concept "person".

After $N = 40$ queries the average precision is 68.93 % and average recall 6.15. As expected, queries had different recall, i.e. results count. Some queries yielded only one result, while others produced over 15 IAPS images ranked according to their semantic similarity with the querying concepts. Precision is calculated for each sequence number in the result set and recall is normalized accordingly. The highest recall ($R = 20$) is for query "animal" because of the repository content and the fact that the most images were already manually tagged with this concept. As can be seen in the next figure, precision and recall fall as the result size increases, as expected. The first result has precision 84.21 %, second 81.58 %, third 65.78 % and so on. Number of currently indexed images is too low for a thorough evaluation of retrieval performance. However, some indicative conclusions can be drawn from these results. The aggregated and averaged results for all $N = 40$ queries are shown in the next figure with query count on x-axis and value range of precision and recall on y-axis. Precision of all queries is indicated with a blue dotted line, while recall is represented with a solid red line.

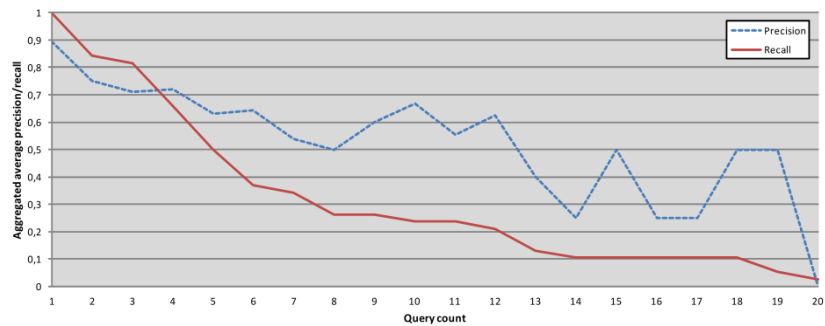

**Fig. 3.** Average retrieval performance after *N* = 40 queries with unit concepts. Average precision is 68.93 % and average recall 6.15. The highest precision is 84.21 % for the first result in sequence.

### 4.1 Discussion

Overall, we found images to be tagged quite exhaustively, which is good for establishing the ground truth in image content, but can be an obstacle for retrieval with semantically similar concepts leading to unrealistically high precision. As can be seen in Fig.3, precision oscillates as the number of results increases, which is a consequence of this over-tagging problem. This problem may be easily resolved if only a randomized subset of tags is used in retrieval.

The second problem is how to define a realistic, useful and at the same time generic measure of semantic relatedness between concepts in ontology. A variety of information-based (corpus dependent) and node distance-based (information independent) similarity measures have different successes in diverse application domains. Many such measures exist and finding the most suitable semantic similarity measure is an ongoing knowledge representation problem.

Another problem is objectivity of tag weight coefficients values. Although potentially a separate source of information valuable for image retrieval, weight values are a product of individual, and subsequently, potentially biased criterion. Tags with unequally distributed weights will produce erroneous relatedness with query tag, especially when coupled with suboptimal semantic distance metrics. However, *kappa* statistic may be used to indicate weights that have inadequate inter-annotator agreement. Tags with such weights should be re-annotated or outlier values left out and weights recalculated. Either way, tag weights have to be consistent across the whole image repository or they will deteriorate the semantic relatedness function. The best way to do this is to establish a clear and formal set of rules for calculating the weights so the influence of personal bias is annulled.

Thirdly, the structure of WordNet's semantic network may be contrary to intuition and too complex [19]. Some subsuming concepts such as "artifact, artifact -- (a man-made object taken as a whole)" or "whole, unit -- (an assemblage of parts that is regarded as a single entity…)" are useful for construction of a well-indented taxonomy, but semantically ambiguous or seemingly redundant to an average person. Intuitively, it seems that a pruned down and lightweight version of WordNet, with new semantic relations such as "Relates-To", "Similar-To" or "Occurs-With", may be better as a knowledge taxonomy for image labeling than the complete WordNet, but without relations that may be lexically ambiguous but could provide fresh and important information content.

WordNet is a huge benefit, but at the same time also a problem. Expressivity of WNtags image annotations is adequate to support content rich and diverse image extraction, but qualitatively the lack of formalism and complex functional relationships between concepts proved limiting and led to ambiguities with different concept senses. For example, a single WordNet concept "snake" has a synonyms "serpent". An average WNtags user expects that semantic distance between all synonyms and another different concept to be the same. However, this is not the case with node distance metrics based on WordNet senses; semantic neighborhoods in WordNet semantic graph around synonyms are not identical. This leads to different image retrieval results when querying these concepts, which may be confusing to the user.

For this reason we noticed that it is unnecessary to observe a wide semantic neighborhood around the search query. Distance of up to 5 or 10 nodes is mostly enough to pick up the majority of concepts related to a query. Larger distances will increase recall, but substantially decrease accuracy and precision in semantic similarity. However, an adaptable algorithm may be envisioned to find a good balance between these two opposing parameters. Such algorithm could start with a small semantic neighborhood around the query and then iteratively increase the distance or include different semantic relations until enough related concepts are found or some other stopping criterion is met. Further research is required to validate this idea.

As mentioned before, it is obvious that only lexically rich knowledge database is not enough for high quality annotation and retrieval. Ontology with a sufficient number of annotating concepts representing objects, events and their various functional relations should allow more accurate image retrieval but also retain a rich annotating glossary which is desirable from the user's perspective.

## 5 Conclusion

WNtags is web-based image annotation tool that may be used to manually identify objects, events and affect that occur in images. Owing to the tool's friendly user interface, new images can easily be added in the document repository and labeled with semantic concepts. WNtags performs image retrieval using WordNet's ontology topology, node distance metrics and collaboratively weighted tags. The results are sorted relative to the query semantic similarity.

The WNtags system was developed as a flexible and modular research tool in areas of annotation and retrieval of real-life emotionally annotated images, but the problems outlined in the previous chapter vividly illustrate difficulties with using WordNet as the knowledge database for image annotation. Only a large ontology deprived of lexical dependencies is able to formally represent the ground truth meaning of image content. For this reason we plan to add SUMO to the WNtags knowledge database. In doing this we would like to retain existing WordNet image labels and map them to SUMO concepts [20]. The implemented concept relatedness metrics is another point which requires further work. A lot of false positive retrieval results are due to inaccurate similarity estimates between query and image concepts. We plan to modularly add different semantic similarity algorithms for information retrieval including information-based metrics.

We would also like to explore modalities of indexing additional multimedia formats apart from pictures, not withstanding specific affective-related domains of static images such as facial expressions and visual stimuli.

## Acknowledgements

This research has been partially supported by the Ministry of Science, Education and Sports of the Republic of Croatia, grant no. 036-0000000-2029. The WNtags tool was developed by student team: Anton Grbin (project leader), Aleksandar Dukovski, Anton Grbin, Jerko Jurin, Dino Milačić, Matija Stepanić, Hrvoje Šimić and Dominik Trupčević as part of their Programming Project course at the University of Zagreb, Faculty of Electrical Engineering and Computing.